%% file: Journal-version-de-base-V2.tex
\documentclass[11pt,a4paper]{article}
\usepackage{amsmath,latexsym,fullpage,amssymb,color}
\usepackage{ifthen,graphics,epsfig}
\usepackage[english]{babel}
\usepackage{times}
\usepackage{float}
\usepackage{xspace}
\usepackage[T1]{fontenc}
\newlength {\squarewidth}

\newtheorem{theorem}{Theorem}
\newtheorem{lemma}{Lemma}
\newtheorem{corollary}{Corollary}

\newcommand{\toto}{xxx}
\newenvironment{proofT}{\noindent{\bf Proof }}
{\hspace*{\fill}$\Box_{Theorem~\ref{\toto}}$\par\vspace{3mm}}

\newenvironment{lemma-repeat}[1]{\begin{trivlist}
\item[\hspace{\labelsep}{\bf\noindent Lemma~\ref{#1} }]}%
{\end{trivlist}}

\newenvironment{theorem-repeat}[1]{\begin{trivlist}
\item[\hspace{\labelsep}{\bf\noindent Theorem~\ref{#1} }]}%
{\end{trivlist}}

\newcounter{linecounter}
\newcommand{\linenumbering}{\ifthenelse{\value{linecounter}<10}
{(\arabic{linecounter})}{(\arabic{linecounter})}}
\renewcommand{\line}[1]{\refstepcounter{linecounter}\label{#1}\linenumbering}
\newcommand{\resetline}[1]{\setcounter{linecounter}{0}#1}
\renewcommand{\thelinecounter}{\ifnum \value{linecounter} > 
9 \else \fi\arabic{linecounter}}
\newfloat{algorithm}{thp}{lop}
\floatname{algorithm}{Algorithm}
\newfloat{construction}{thp}{lop}
\floatname{construction}{Construction}
\newcommand{\Xomit}[1]{}

\newcommand{\CONS}{\mathit{CS}}
\newcommand{\ISNAP}{\mathit{IS}}

\newcommand{\REG}{\mathit{REG}}

\newcommand{\VIEW}{\mathit{VIEW}}

\newcommand{\return}{{\sf return}}
\renewcommand{\max}{{\sf max}}
\renewcommand{\min}{{\sf min}}
\newcommand{\wait}{{\sf wait}}
\newcommand{\CARW}{{\cal CARW}_{n,t}}
\newcommand{\snapshot}{{\sf snapshot}}
\newcommand{\wwrite}{{\sf write}}

\newcommand{\writesnap}{{\sf write\_snapshot}}
\newcommand{\propose}{{\sf propose}}

\title{{\bf $t$-Resilient $k$-Immediate Snapshot and\\
           its Relation with Agreement Problems}\thanks{Parts of 
    this work were presented at  the 23rd International Colloquium on Structural
    Information and Communication  Complexity (SIROCCO'2016)~\cite{DFRR16}
    and at the 22nd International Symposium on Stabilization, Safety, and Security of Distributed Systems (SSS'2020).}
}

\author{Carole Delporte$^{\dag}$,
        Hugues Fauconnier$^{\dag}$,
        Sergio Rajsbaum$^{\circ}$, 
       Michel Raynal$^{\star,\ddag}$\\~\\
$^{\dag}$  IRIF,  Universit\'e Paris Diderot, Paris, France \\
$^{\circ}$ Instituto de Matem\'aticas, UNAM, M\'exico D.F, 04510, M\'exico\\
$^{\star}$ Institut Universitaire de France\\
$^{\ddag}$ IRISA, Universit\'e de Rennes, 35042 Rennes, France 
}

\begin{document}
\maketitle
\setcounter{footnote}{0}


\begin{abstract}
An immediate snapshot object is a high level communication object,
built on top of a read/write distributed system in which all except
one processes may crash. It provides the processes with a single
operation denoted $\writesnap_k()$, which allows a process to write a
value and obtain a set of pairs (process id, value) satisfying some
set containment properties, that represent a snapshot of the values
written to the object, occurring immediately after the write step.

Considering an $n$-process model in which up to $t$ processes may
crash, this paper introduces first the $k$-resilient immediate snapshot object, 
which is a natural generalization  of the basic immediate snapshot
(which corresponds to the case $k=t=n-1$). 
In addition to the set containment properties of the basic immediate snapshot,
a $k$-resilient immediate snapshot object requires that each set 
returned to a process contains at least $(n-k)$ pairs. 

The paper  first shows  that, for $k,t<n-1$,  $k$-resilient immediate 
snapshot is impossible in asynchronous read/write systems. 
Then the paper investigates a model of computation where the processes
communicate with each other by accessing $k$-immediate
snapshot objects, and shows that this model is stronger than
the $t$-crash model.
Considering the space of $x$-set agreement problems (which are
impossible to solve in systems such that $x\leq t$), the paper shows
then that $x$-set agreement can be solved in read/write systems
enriched with $k$-immediate snapshot objects for
$x=\max(1,t+k-(n-2))$.  It also shows that, in these systems,
$k$-resilient immediate snapshot and consensus are equivalent when
$1\leq t<n/2$ and $t\leq k\leq (n-1)-t$.  Hence,
the paper establishes strong relations linking fundamental distributed
computing objects (one related to communication, the other to
agreement), which are impossible to solve in pure read/write systems.

~
\end{abstract}

\newpage

\section{Introduction}

\paragraph{Context}
This article considers the $t$-crash model consisting of $n$
asynchronous processes, among which any subset of at most $ t$
processes may crash, and communicate through a shared memory composed
of single writer/multi reader (SWMR) atomic registers. The $(n-1)$-crash model is also called wait-free model~\cite{H91}. We keep the
term $t$-resilience for algorithms. Several progress conditions have
been associated with $(n-1)$-resilient algorithms:
wait-freedom~\cite{H91}, non-blocking~\cite{HW90}, or
obstruction-freedom~\cite{HLM03} (see a unified presentation in Chapter
5 of~\cite{R13}). This article focuses on the wait-free condition, in
the context of tasks: every non-failed process has to produce an
output value. A task is defined in terms of (a) possible inputs to the
processes, and (b) valid outputs for each assignment of input values
(tasks are precisely defined in~\cite{BGLR01,HRR13,HS99}).  Of special
importance is the family of $x$-set agreement tasks~\cite{C93}, one
for each integer value of$ x$, $1 \leq x \leq n$.  Set agreement was
introduced to show a hierarchy of tasks whose solvability depends on
$t$, the number of processes that may crash.  In the $x$-set agreement
task, processes decide at most $x$ different values, out of their
input assignments. When $x = 1$, $x$-set agreement is the celebrated
consensus task, which is impossible even in the presence of a single
process crash~\cite{FLP85,LA87}. More generally, $x$-set agreement is
solvable if and only if $t < x$, a result proved using algebraic
topology~\cite{BG93,BGLR01,HS99,SZ00}.  There are characterizations of
the solvability of any given task, in the $t$-crash model, and in
others (for an overview of results see~\cite{HKR14}).

\paragraph{Immediate snapshot object}
The {\it immediate snapshot} (IS) communication object was first introduced 
in~\cite{BG93-b,SZ00}, and then further investigated as an ``object''
in~\cite{BG93}. This object is at the heart of the {\it iterated immediate 
snapshot} (IIS) model introduced in~\cite{BG97,HS94}, which 
consists of $n$  asynchronous processes, among which any subset at most  
$(n-1)$ processes may crash.
These processes  execute a sequence of asynchronous rounds, and
each round is provided with  exactly one IS object,
which allows the processes to communicate only during this round. 
More precisely,  for any $x>0$, a process accesses  the $x$-th immediate 
snapshot only when it executes the $x$-th round, and it accesses it only 
once.

From an abstract point of view, an IS object $\ISNAP$, can be seen as
an initially empty set, which can then contain up to $n$ pairs (one
per process), each made up of a process index and a value.  This
object provides each process with a single operation denoted
$\writesnap()$, that it can invoke only once.  The invocation
$\ISNAP.\writesnap(v)$ by a process $p_i$ adds the pair $\langle
i,v\rangle$ to $\ISNAP$ and returns a set of pairs belonging to
$\ISNAP$ such that the sets returned to the processes that invoke
$\writesnap()$ satisfy specific inclusion properties.  It is important
to notice that, in the IIS model, the processes access the sequence of
IS objects one after the other, in the same order, and asynchronously.

\paragraph{Contribution of the paper}
%
As previously said, the IS object was designed for the wait-free model
(i.e., $t=n-1$).  This paper considers it in the context of the
$t$-crash $n$-process system models where $t<n-1$.  To this end it
generalizes the IS object by introducing the notion of a $k$-immediate
snapshot ($k$-IS) object. Such an object provides the processes with a
single operation denoted $\writesnap_k()$ which, in addition to the
properties of an IS object, returns a set including at least $(n-k)$ pairs.
Hence, for $k<n-1$, due to the implicit synchronization implied by the 
constraint on the minimal size of  the sets it returns, a $k$-IS object 
allows processes to obtain more information from the whole set of processes
than a simple IS object (which may return  sets containing 
less than $(n-k)$  pairs).

The obvious question is then the implementability of a $k$-IS object
in the $t$-crash $n$-process asynchronous read/write model.  
The paper shows first that, differently from the basic 
IS object which can be implemented in the wait-free model, no $k$-IS object
where $k<n-1$, can  be implemented in a $1$-crash $n$-process read/write system.

This impossibility result is far from being the first impossibility
result in the presence of asynchrony and process crashes.  The most
famous of them, which  concern agreement problems, 
are the impossibility of Consensus (CONS) in the
presence of even a single process crash~\cite{FLP85,LA87}, and the
impossibility of $x$-set agreement ($x$-SA) when 
$x\leq t$~\cite{BG93-b,HS99,SZ00} (let us remind that CONS is $1$-SA). These 
objects are at the heart of the theory of fault-tolerant distributed computing.

Hence, a second natural question: 
Are they relations linking the previous ``impossible'' objects, 
namely $k$-IS and $x$-SA? 
The paper provides the following answers to this question.
\begin{itemize}
\vspace{-0.2cm}
\item
Let $1\leq k \leq t < n$. It is possible to implemented a $k$-IS object 
in a $t$-crash $n$-process read/write system enriched with consensus objects. 
\vspace{-0.2cm}
\item
Let $1 \leq t <n/2$ and $t\leq k \leq (n-1)-t$. 
$k$-IS and Consensus are equivalent in a $t$-crash $n$-process read/write 
system. ($A$ and $B$ are equivalent if $A$ can be implemented in the
$t$-crash $n$-process read/write system enriched with $B$, and reciprocally.)
\vspace{-0.2cm}
\item
Let $(n-1)/2 \leq k \leq n-1$ and $ (n-1)-k \leq t \leq k$. 
 It is possible to implemented an $x$-SA object, where $x=t+k-(n-2)$,  
in a $t$-crash $n$-process read/write system enriched with $k$-IS objects.
\end{itemize}

\Xomit{
At first glance, this impossibility result may seem surprising. 
An IS object is a snapshot object (a) whose operations $\wwrite()$ 
and $\snapshot()$ are glued together in a single operation  $\writesnap()$, 
and (b) satisfying an additional property  linking the sets of pairs 
returned by  concurrent invocations (called {\it Immediacy} property, 
Section~\ref{sec:one-shot-immediate-snapshot}). 
Then, as already indicated, 
a $t$-IS object  is an IS object such that the sets returned by  
$\writesnap()$ contain  at least $(n-t)$ pairs ({\it Output size} property,
Section~\ref{sec:k-immediate-snapshot}). 
The same Output size property on the sets 
returned by a snapshot object can be trivially implemented in  a $t$-crash 
$n$-process model. Let us call $t$-snapshot such a  constrained 
snapshot object. Hence, while a $t$-snapshot object can be implemented in 
the $t$-crash $n$-process model, a $t$-IS object cannot when $0<t<n-1$. 
}

\paragraph{Roadmap}

The paper develops the previous map. It is made up of \ref{sec:conclusion} 
sections. 
Section~\ref{sec:model-definitions} presents the basic $t$-crash $n$-process
asynchronous read/write model, and the definitions of the IS
and $x$-SA objects. Section~\ref{sec:k-SA-def-impossibility} defines the 
$k$-IS object and its impossibility in the previous basic model.

The other sections are on the power of $k$-IS with respect to $x$-SA. 
Section~\ref{sec:k-IS-to-x-SA} shows that $x$-SA can be built in the $t$-crash 
$n$-process asynchronous read/write model enriched with $k$-IS objects, 
for $x=\max(1,t+k-(n-2))$. 
Section~\ref{sec:equiv-to-cons} shows that $t$-IS and CONS  are equivalent 
in the $t$-crash $n$-process asynchronous read/write model when $1\leq t< n/2$.
Section~\ref{sec:cons-to-k-IS} shows that CONS is stronger than $k$-IS
when $n/2 \leq t \leq k  <n-1$. 
Finally, Section~\ref{sec:conclusion} concludes the paper. 
An illustration of the previous results is presented in 
Table~\ref{fig:matrix-example}, which considers a system of $n=11$ processes.

\begin{table}[th]
\centering{

\begin{center}
\renewcommand{\baselinestretch}{1}
\small
\begin{tabular}{|c||c|c|c|c|c|c|c|c|c|c|c|}
\hline
   $k\rightarrow$  
     & 1  & 2  & 3  & ..  & ..  & ..  & $n-4$  & $n-3$  & $n-2$  & $n-1$ \\
   $t \downarrow$ & 1  & 2  & 3  & 4  & 5  & 6  & 7  & 8  & 9  & 10  \\ 
\hline\hline
 1
& 1-SA & 1-SA & 1-SA & 1-SA & 1-SA  & 1-SA & 1-SA & 1-SA & 1-SA & 2-SA \\
\hline 
2
&  & 1-SA & 1-SA & 1-SA & 1-SA  & 1-SA & 1-SA & 1-SA & 2-SA & 3-SA \\
\hline 
3
& & & 1-SA & 1-SA & 1-SA & 1-SA  & 1-SA & 2-SA & 3-SA & 4-SA  \\
\hline 
4
& & & & 1-SA & 1-SA  & 1-SA & 2-SA & 3-SA & 4-SA & 5-SA \\ 
\hline 
$5<n/2$
& & & & & 1-SA & 2-SA & 3-SA & 4-SA & 5-SA &6-SA \\ 
\hline 
$6\geq n/2$
& & & & & & 3-SA & 4-SA & 5-SA & 6-SA & 7-SA\\ 
\hline 
$7=n-4$
& & & & & & & 5-SA &  6-SA & 7-SA & 8-SA \\ 
\hline 
$8=n-3$
& & & & & & &  & 7-SA & 8-SA & 9-SA \\ 
\hline 
$9=n-2$
& & & & & & &  &  & 9-SA & 10-SA\\ 
\hline 
$10=n-1$
& & & & & & &  &  &  & 11-SA\\ 
\hline 
\end{tabular}
\end{center}
\vspace{-0.4cm}
\caption{From $k$-IS to $x$-SA with $x=\max(1,t+k-(n-2))$  ($n=11$)}
\label{fig:matrix-example}
}
\end{table}

\section{Basic Model, Immediate Snapshot, and $x$-Set Agreement}
\label{sec:model-definitions}

\subsection{Basic read/write system model}

\paragraph{Processes}
\label{sec:process-model}
The computing model is composed of a set of $n\geq 3$ sequential processes
denoted $p_1$, ..., $p_n$. Each process is asynchronous which means
that it proceeds at its own speed, which can be arbitrary and remains
always unknown to the other processes.  

A process may halt prematurely (crash failure), but executes correctly
its local algorithm until it possibly crashes. The model parameter $t$
denotes the maximal number of processes that may crash in a  run.
A process that crashes in a run is said to be {\it faulty}. Otherwise, 
it is {\it correct} or {\it non-faulty}. Let us notice that, as 
a faulty process behaves correctly until it  crashes, no process knows 
if it is correct or faulty. Moreover, due to process asynchrony, 
no process can know if another process crashed or is only very slow.

It is assumed that (a) $0<t<n$ (at least one process may crash and 
at least one process does not crash), and  
(b) any process, until it possibly crashes,  executes correctly 
the algorithm assigned to it.

\paragraph{Communication layer}
The processes cooperate by reading and writing 
Single-Writer Multi-Reader (SWMR) atomic read/write registers~\cite{L86}. 
This means that the shared memory can be seen as a set of  arrays 
$A[1..n]$ where, while $A[i]$ can be read by all processes, 
it can be written only by $p_i$.

\paragraph{Notation} 
The previous  model
is denoted $\CARW[\emptyset]$ (which means ``Crash Asynchronous Read/Write
with $n$ processes, among which up to $t$ may crash'').
A model constrained by a predicate on $t$ (e.g. $t<a$) 
is denoted  $\CARW[t<a]$.
It is assumed that  at least one process does not crash, 
$\CARW[t=n-1]$ is a synonym of $\CARW[\emptyset]$, 
which (as always indicated) is called {\it wait-free} model. 
When considering $t$-crash models, 
$\CARW[t <a]$ is less constrained than $\CARW[t<a-1]$.
More generally, $\CARW[P,T]$ denotes the system model 
$\CARW[\emptyset]$ restricted by the predicate $P$, and enriched with any 
number of shared objects of the type $T$ (e.g., consensus objects).

Shared objects are denoted with capital letters. 
The local  variables of a process $p_i$ are denoted with lower case letters,
sometimes suffixed by the process index $i$.


\subsection{Immediate snapshot}
\label{sec:one-shot-immediate-snapshot}
The immediate snapshot (IS) object~\cite{BG93} was informally presented 
in the introduction. Defined in the context of the wait-free
model (i.e., $t=n-1$), it can be seen as a variant of the snapshot object
introduced in~\cite{AADGMS93,A94}. While a snapshot object provides 
the processes with two operations ($\wwrite()$ and $\snapshot()$)
which can be invoked separately by a process  (usually a process invokes 
$\wwrite()$  before $\snapshot()$), a one-shot immediate snapshot object 
provides the processes  with a single operation $\writesnap()$ (one-shot 
means that a process may  invoke $\writesnap()$ at most once). 

\paragraph{Definition}
Let $\ISNAP$  be an IS object. It is a set, initially empty, 
that will contain pairs made up of a process index and a  value. 
Let us consider a process $p_i$ that invokes 
$\ISNAP.\writesnap(v)$. This invocation adds the pair 
$\langle i,v\rangle$ to $\ISNAP$ (contribution of $p_i$ to  $\ISNAP$), 
and returns to $p_i$ a set,  called  view and denoted $view_i$,
such that the sets returned to the processes 
collectively satisfy the following properties. 
\begin{itemize}
\vspace{-0.1cm}
\item Termination. The invocation of  $\writesnap()$ by a correct process
                   terminates. 
\vspace{-0.2cm}
\item Self-inclusion. $\forall~i:~ \langle i,v\rangle \in view_i$. 

\vspace{-0.2cm}
\item Validity.  $\forall~i: (\langle j,v\rangle \in view_i)\Rightarrow$
                 $p_j$ invoked $\writesnap(v)$. 
\vspace{-0.2cm}
\item Containment. 
$\forall~i,j:~ (view_i\subseteq view_j)\vee(view_j\subseteq view_i)$. 
\vspace{-0.2cm}
\item Immediacy. 
  $\forall~i,j:~ (\langle i,v \rangle \in view_j)\Rightarrow 
                  (view_i \subseteq view_j)$.\footnote{An equivalent 
formulation of the Immediacy property is: 
 $\forall~i,j:~ \big((\langle i,- \rangle \in view_j)
           \wedge(\langle j,- \rangle \in view_i)\big)
           \Rightarrow  (view_i = view_j)$.} 
\end{itemize}

Implementations of an IS object in  the wait-free model $\CARW[t=n-1]$ 
are described in~\cite{BG93,GR10,RR13,R13}. 
While both a one-shot snapshot object and an IS object 
satisfy the Self-inclusion, Validity and Containment properties, 
only an IS object satisfies the Immediacy property. This additional 
property creates an important difference, from which follows that, 
while a snapshot object is atomic (operations on a snapshot object
can be linearized~\cite{HW90}), an IS object is not atomic (its operations 
cannot always be linearized). 
However, an IS object is set-linearizable 
(set-linearizability allows several operations to be linearized 
at the same point of the time line~\cite{CRR15,N94}). 

\Xomit{
\paragraph{The iterated immediate snapshot {(IIS)}  model}
This model (introduced in~\cite{BG97}) considers $t=n-1$. Its shared
memory is composed of a (possibly infinite) sequence of IS objects:
$\ISNAP[1]$, $\ISNAP[2]$, ..., which  are accessed sequentially
and asynchronously by the processes according to the following
round-based pattern executed by each process $p_i$.  The variable
$r_i$ is local to $p_i$; it denotes its current round number.
\vspace{-0.2cm}

{\centering
\begin{tabbing}
aaaaaaaaaa\=aa\=aa\=aa\=aaa\=aaa\=\kill
\>$r_i\leftarrow 0$;  $\ell s_i \leftarrow$  initial local state of $p_i$ 
  (including its input, if any);\\
\>{\bf repeat forever} \%  asynchronous IS-based  rounds\\
\>\>  $r_i \leftarrow r_i +1$;\\
\>\>  $view_i \leftarrow \ISNAP[r_i].\writesnap(\ell s_i)$;\\
\>\> computation of a new local state $\ell s_i$ 
       (which contains $view_i$)\\
\>{\bf end repeat}.
\end{tabbing}
}
\vspace{-0.2cm}
\noindent
As indicated in the Introduction, when considering distributed tasks (as 
formally defined in~\cite{BGLR01,HRR13,HS99}), the IIS model and 
$\CARW[t=n-1]$  have the same computational power~\cite{BG97}. 
} 

\subsection{$x$-Set agreement}
$x$-Set agreement was introduced by S. Chaudhuri~\cite{C93}
to investigate the relation linking the number $x$ of different values 
that can be decided in an agreement problem, and the maximal number of faulty 
processes $t$. 
It generalizes consensus which corresponds to the instance $x=1$. 

An $x$-set agreement ($x$-SA) object is a one-shot object 
that provides the processes with a single operation denoted 
$\propose_x()$. This operation allows the invoking  process $p_i$
to propose a value, which is called {\it proposed} value, an is 
passed as an input parameter. It returns a value, called {\it decided} value. 
The object is defined by the following set of properties. 
\begin{itemize}
\vspace{-0.1cm}
\item Termination. The invocation of  $\propose_x()$  by a correct process
                   terminates.
\vspace{-0.2cm}
\item Validity. A decided value is a proposed value. 
\vspace{-0.2cm}
\item Agreement. No more than $x$ different values are decided. 
\end{itemize}
It is shown in~\cite{BG93-b,HS99,SZ00} that the problem is 
impossible to solve in  $\CARW[x\leq t]$.

\section{$k$-Immediate Snapshot and its $t$-Resilience Impossibility}
\label{sec:k-SA-def-impossibility}

\subsection{Definition and a property of $k$-immediate snapshot}
A $k$-immediate snapshot ($k$-IS) object is an 
immediate snapshot object with the following additional  property. 
\begin{itemize}
\vspace{-0.1cm}
\item Output size. 
The set $view$ obtained by a process is such that $|view|\geq n-k$.
\end{itemize}

This means that in addition to the Self-inclusion, Validity, Containment,
and Immediacy properties, the set returned to a process contains at least 
$(n-k)$ pairs. 
The associated operation is denoted $\writesnap_k()$. 

\paragraph{$k$-Immediate snapshot vs $x$-set agreement}
When considering  a $k$-IS object and a $x$-SA object, 
we have the following differences. 
\begin{itemize}
\vspace{-0.2cm}
\item On concurrency. 
An $x$-SA object is atomic (linearizable), while 
a $k$-SA object is not (it is only set-linearizable~\cite{CRR15,N94}). 
In other words,  $k$-SA objects ``accept'' concurrent accesses
(this is captured by the Immediacy property),  while  $x$-SA objects do not. 
\vspace{-0.2cm}
\item On the values returned. 
When considering an $x$-SA object, 
each process $p_i$  knows that each other process $p_j$ (which returns from its 
invocation of $\propose_x()$) obtains a single value, but 
it does know which one (uncertainty); $p_i$  knows only that at most $k$
values are decided by all  processes (certainty). 

When considering a $k$-IS object, 
each process $p_i$ knows that each other process $p_j$ (which returns from its 
invocation of $\writesnap_k()$) obtains a set of pairs $view_j$ 
that is included in, is  equal to, or includes its own set of pairs
(certainty due to the containment property), but it does not know 
the  size of $view_j$  (uncertainty).
\end{itemize}

\paragraph{A property associated with $k$-IS objects}
The next theorem characterizes the power of a  $k$-IS object in term
of its Output size and  Containment properties. 

\begin{theorem}
\label{theo:main-property-k-IS}
Let us consider a  $k$-{\em IS} object,  and 
assume that all correct processes invoke $\writesnap_k()$.
If  the size of the smallest view obtained by a process is $\ell$
($\ell\geq n-k$), there is a set $S$ of processes such that 
$|S|=\ell$ and each process of $S$ obtains the smallest view 
or crashes during its invocation of  $\writesnap_k()$.
\end{theorem}

\begin{proofT}
It follows from the Output size property of the $k$-IS object that no
view contains less than $\ell \geq n-k$ pairs.  Let $min\_view$ be the 
smallest view returned by a process; hence $\ell=|min\_view|$.  

Let us consider a process $p_i$ such that $(\langle i,-\rangle\in min\_view)$, 
which returns a view.   Due to 
(a) the Immediacy property (namely $(\langle i,-\rangle\in min\_view)$ 
$\Rightarrow$ $(view_i\subseteq min\_view)$) and 
(b) the minimality of $min\_view$, it follows that $view_i=min\_view$.  
As this  is true for each process whose pair participates in $min\_view$, 
it follows that there is a set $S$ of processes such that 
$|S|=\ell\geq n-k$,  and  each of these  processes obtains  
$min\_view$, or crashes during its invocation of $\writesnap_k()$.
Due to the Containment property, the others processes crash  or  
obtain views which are a superset of $min\_view$.
\renewcommand{\toto}{theo:main-property-k-IS}
\end{proofT}

This theorem establishes the most important property of a $k$-IS object. 
This property is used in nearly all lemmas and theorems appearing in the paper.

\subsection{An impossibility result}

\begin{theorem}
\label{theo:basic-imposs-k-IS}
A {\em $k$-IS} object cannot be implemented in $\CARW[k < t]$.
\end{theorem}

\begin{proofT}
To satisfy the output size property, the view obtained by a process $p_i$ 
must contain pairs from $(n-k)$ different processes.  
If $t$ processes crash (e.g.,  initial crashes), a process can obtain at most  
$(n-t)$ pairs. If $t>k$, we have $n-t<n-k$. It follows that, after it has 
obtained pairs from $(n-t)$ processes,  a process can remain blocked forever 
waiting for the $(t-k)$ missing pairs. 
\renewcommand{\toto}{theo:basic-imposs-k-IS}
\end{proofT}

\begin{theorem}
\label{theo:k-IS-t-crashes} 
Let $k < n-1$. It is impossible to 
implement a {\em $k$-IS} object in  $\CARW[1 \leq t\leq k <n-1]$.
\end{theorem}

\begin{proofT}
The case where $k<t$ was proved in Theorem~\ref{theo:basic-imposs-k-IS}.
Hence, the proof  considers the case $1=t\leq k<n-1$ (this constraint 
explains the model assumption $n\geq 3$, Section~\ref{sec:process-model}).   
If, for $k\leq n-1$, there is no implementation of a $k$-IS object 
in $\CARW [t=1]$, there is no implementation either for $t\geq 1$. 
The proof is by contradiction, namely,  assuming an implementation of a  
$k$-IS object, where $k < n-1$,  in $\CARW[t=1]$, we show that
it is possible to solve consensus in $\CARW[t=1,\mbox{$k$-IS}]$.
As consensus cannot be solved in $\CARW[t=1]$, it follows that 
$k$-IS cannot be implemented in $\CARW[1\leq t\leq k]$.

Let us recall the main property of $k$-IS (captured by 
Theorem~\ref{theo:main-property-k-IS}). Let $\ell$ be the size of 
the smallest view ($min\_view$) returned by a process. 
There is a set $S$ of $\ell$  processes such that 
any process of $S$ returns $min\_view$ or crashes, and $\ell \geq n-k$. 
As $k<n-1$ (theorem assumption), we have $\ell \geq 2$, 
which means that at least two processes obtain $min\_view$. 
It follows that, if a process obtains the views returned 
by the $k$-IS object to $(n-1)$ processes, one of these views is 
necessarily $min\_view$. This constitutes Observation $O$. 

\begin{algorithm}[h!]
\centering{\fbox{
\begin{minipage}[t]{150mm}
\footnotesize 
\renewcommand{\baselinestretch}{2.5}
\resetline
\begin{tabbing}
aaaaa\=aaa\=aaaaa\=aaaaaa\=\kill

{\bf operation} $\propose_1(v)$ {\bf is}\\

\line{Reduction-01} 
\>  $view_i \leftarrow \ISNAP.\writesnap_k(v)$;\\

\line{Reduction-02}
\>    $VIEW[i]\leftarrow view_i$;\\

\line{Reduction-03}
\>  $\wait(|\{~j\mbox{ such that } VIEW[j]\neq\bot\}|=n-t)$;\\

\line{Reduction-04}
\>  
{\bf let} $view$ {\bf be} the smallest of  the previous $(n-t)$ views;\\

\line{Reduction-05}
\>  $\return(\mbox{smallest proposed value in }view)$\\

{\bf end operation}.
\end{tabbing}
\end{minipage}
}
\caption{
Solving consensus in $\CARW[ t=1,\mbox{$k$-IS}]$ 
(code for $p_i$)}
\label{algo:reduction-cons-to-k-IS}
}
\end{algorithm}

Let us now consider Algorithm~\ref{algo:reduction-cons-to-k-IS}. 
In addition to a $k$-IS object denoted $\ISNAP$, the processes access
an array $\VIEW[1..n]$ of SWMR atomic registers, initialized to
$[\bot,\cdots,\bot]$.  The aim of $\VIEW[i]$ is to store the view
obtained by $p_i$ from the $k$-IS object $\ISNAP$.
When it calls $\propose_1(v)$, a process $p_i$ invokes first the
$k$-IS object, in which it deposits the pair $\langle i,v\rangle$, and
obtains a view from it (line~\ref{Reduction-01}), that it writes in
$\VIEW[i]$ to make it publicly known (line~\ref{Reduction-02}).  Then,
it waits until it sees the views of at least $(n-1)$ processes
(line~\ref{Reduction-03}). Finally, $p_i$ extracts from these views
the one with the smallest cardinality (line~\ref{Reduction-04}), and returns 
the smallest value contained in this smallest view (line~\ref{Reduction-05}).

We show that this reduction algorithm solves consensus in
$\CARW[t=1,\mbox{$k$-IS}]$.  As at least $(n-1)$ processes do not
crash, and write in their entry of the array $\VIEW[1..n]$, no correct
process can block forever at line~\ref{Reduction-02}, proving the
Termination property of consensus.

As $\ell\geq n-k \geq 2$, it follows from Observation $O$  
that at least one of the views  obtained by a process at 
line~\ref{Reduction-03} is necessarily $min\_view$.
It follows that each process that executes line~\ref{Reduction-03}
obtains $min\_view$ and returns  its smallest value at  
line~\ref{Reduction-04}), proving the Agreement property of consensus. 

The consensus Validity property follows directly from $k$-IS Validity
property, and the observation that any set $view$ contains only
proposed values line~\ref{Reduction-04}).
\renewcommand{\toto}{theo:k-IS-t-crashes}
\end{proofT}

\paragraph{Remark}
When considering the algorithm described in 
Figure~\ref{algo:reduction-cons-to-k-IS}, let us 
observe that, as $n-k \leq n-t$, the array $\VIEW[1..n]$ 
can replaced by a second  $k$-immediate snapshot object $\ISNAP2$.
We obtain then the following algorithm.

{\footnotesize 
\renewcommand{\baselinestretch}{2.5}
\begin{tabbing}
aaaaa\=\kill

$~~~~~~~~~~~~~~~~~~~~~~~~~~~~~~~~~~~~~~~~~~~~~$ 
{\bf operation} \= $\propose_1(v)$ {\bf is}\\
 
\>  $view1_i \leftarrow \ISNAP.\writesnap_k(v)$;\\

\>   $view2_i \leftarrow \ISNAP2.\writesnap_k(view1_i)$;\\
\>  
{\bf let} $view$ {\bf be} the smallest view in $view2_i$;\\

\>  $\return(\mbox{smallest proposed value in }view)$\\

$~~~~~~~~~~~~~~~~~~~~~~~~~~~~~~~~~~~~~~~~~~~~~$ 
{\bf end operation}.
\end{tabbing}
}

\section{From $k$-Immediate Snapshot to $x$-Set Agreement}
\label{sec:k-IS-to-x-SA}

This section proves the content of Table~\ref{fig:matrix-example}, 
namely $x$-SA can be implemented in the system model $\CARW[t\leq k<n-1]$, 
for  $x=\max(1,t+k-(n-2))$.
Interestingly, the algorithm providing such an implementation 
is Algorithm~\ref{algo:reduction-cons-to-k-IS}, whose operation 
name is now  $\propose_x()$ (instead of $\propose_1(v)$).

\begin{theorem}
\label{theo:cartography}
Let  $x=\max(1,k+t-(n-2))$. 
Algorithm~{\em\ref{algo:reduction-cons-to-k-IS}} implements an  {\em $x$-SA}
object in $\CARW[1\leq t \leq k <n-1,\mbox{\em $k$-IS}]$.
\end{theorem}

\begin{proofT}
The consensus Termination follows directly from the Termination property of the 
underlying $k$-IS object $\ISNAP$, the fact that there are at least $(n-t)$ 
correct processes, and the assumption that  all correct processes  invoke   
$\propose_x()$. The consensus Validity property follows directly from the 
Validity property of the $\ISNAP$. 

As far as the consensus Agreement property is concerned, we have the following. 
Due to Theorem~\ref{theo:main-property-k-IS}, a set of 
$\ell\geq n-k$ processes obtain the smallest possible 
view $min\_view$, which is such that $|min\_view|=\ell \geq n-k$. 
It follows that, at most $k$ processes obtain a view different 
from $min\_view$. In the worst case, these $k$ views are different.
Consequently, there are at most $k+1$ different views,  namely
$min\_view$, $V(1)$, ..., $V(k)$, and due to their Containment property, 
we have $min\_view \subset V(1)\subset \cdots\subset V(k)$. 
The rest of the proof is a case analysis  according to the value 
of $(n-t)$ with respect to $k$. 
\begin{itemize}
\vspace{-0.2cm}
\item $n-t>k$. 
In this case, a process obtains at line~\ref{Reduction-03}
views from $(n-t)$ processes, and in the first case 
it obtains the views $V(1)$, ..., $V(k)$. But as $n-t>k$ 
it also obtains $min\_view$ from at least one process. 
It follows that, all processes see $min\_view$,  and they consequently decide 
the same value at line~\ref{Reduction-05}. Hence, $(n-t>k)\Rightarrow (x=1)$. 
\vspace{-0.2cm}
\item $n-t=k$. 
In this case, it is possible that some  processes do not obtain $min\_view$ 
at line~\ref{Reduction-03}. But, if this occurs,  they necessarily  obtain 
the views from the $n-t=k$ processes  that deposited  $V(1)$, ..., $V(k)$ 
in $\VIEW[1..n]$.  Hence, all these processes obtains $V(1)$ at 
line~\ref{Reduction-03}, and decide  consequently the same value  from $V(1)$. 
As the decided values are decided from the views $\min\_view$ and $V(1)$, 
we have $(n-t=k)\Rightarrow (x=2)$. 
\vspace{-0.2cm}
\item 
$n-t=k-1$. 
In this case, it is possible that, at line~\ref{Reduction-03},
some  processes do not obtain not only $min\_view$, but also $V(1)$
and decide the smallest value of $V(2)$. 
As the decided values are then decided from the views $\min\_view$, 
 $V(1)$,  and $V(2)$,  we have $(n-t=k-1)\Rightarrow (x=3)$. 
\vspace{-0.2cm}
\item 
Applying the same reasoning to the general case 
 $n-t=k-c$, we obtain  $(n-t=k-c)\Rightarrow (x=2+c)$. 
\end{itemize}
Abstracting the previous case analysis, we obtain 
$x=1$ (consensus) for $n-t>k$, and $x=k+t-(n-2)$, i.e., when  $n-t=k-x+2$, 
from which follows that $x=\max(1,k+t-(n-2))$, 
which completes the proof of the theorem.
\renewcommand{\toto}{theo:cartography}
\end{proofT}

\noindent
The next corollary is  a re-statement of Theorem~\ref{theo:cartography}
for $x=1$. 
\begin{corollary}
\label{coro:cartography}
Algorithm~{\em\ref{algo:reduction-cons-to-k-IS}} implements a consensus
object in $\CARW[1\leq t <n/2, t\leq k\leq (n-1)-t,\mbox{\em $k$-IS}]$.
\end{corollary}

\section{An Equivalence Between $k$-Immediate Snapshot and Consensus}
\label{sec:equiv-to-cons}

This section shows first that consensus is strong enough to implement 
a $k$-IS object when $t\leq k$. Combining this result 
with the fact consensus can be implemented from a $k$-IS object
in $\CARW[1\leq t<n/2, t\leq k\leq (n-1)-t]$
 (Corollary~\ref{coro:cartography}), we obtain that consensus and $k$-IS  
are equivalent in $\CARW[1\leq t<n/2, t\leq k\leq (n-1)-t]$.

\subsection{From consensus to $k$-IS in $\CARW[t\leq k\leq n-1]$} 
Algorithm~\ref{algo:reduction-k-IS-to-cons} describes 
a reduction of $k$-IS to consensus in  $\CARW[0<t\leq k \leq n-1]$. 
This algorithm uses three shared data structures. 
The first is an  array $\REG[1..n]$ of SWMR atomic registers (where 
$\REG[i]$ is associated with $p_i$),
the second is a consensus objects denoted $\CONS$, and the third
is an immediate snapshot object denoted $\ISNAP$
(let us recall that such an object can be implemented in $\CARW[t \leq n-1]$).

\begin{algorithm}[h!]
\centering{\fbox{
\begin{minipage}[t]{150mm}
\footnotesize 
\renewcommand{\baselinestretch}{2.5}
\resetline
\begin{tabbing}
aaaaa\=aaa\=aaaaa\=aaaaaa\=\kill

{\bf operation} $\writesnap_k(v_i)$ {\bf is}\\

\line{Reduction-Cons-01} 
\>  $\REG[i] \leftarrow  v_i$;\\

\line{Reduction-Cons-02}
\> 
${\sf wait}$ \big($|j \mbox{ such that } \REG[j]\neq\bot\}|\geq n-k$\big);\\

\line{Reduction-Cons-03}
\>
$aux_i\leftarrow
   \{\langle j,\REG[j] \rangle \mbox{ such that } \REG[j]\neq\bot\}$;\\

\line{Reduction-Cons-04}
\> $view_i \leftarrow \CONS.\propose_1(aux_i)$;\\

\line{Reduction-Cons-05}
\> {\bf if} \= $(\langle i,v_i \rangle \in view_i)$\\

\line{Reduction-Cons-06}
\>\> {\bf then} \= $\return(view_i)$\\

\line{Reduction-Cons-07}
\>\> {\bf else} \> $aux_i \leftarrow \ISNAP.\writesnap(v_i)$;\\

\line{Reduction-Cons-08}
\>\>\>  $view_i \leftarrow view_i \cup aux_i$; \\

\line{Reduction-Cons-09}
\> \>\> $\return(view_i)$\\

\line{Reduction-Cons-10}
\> {\bf end if} \\

{\bf end operation}.
\end{tabbing}
\end{minipage}
}
\caption{Implementing $k$-IS in $\CARW[0<t\leq k\leq n-1,\mbox{CONS}]$
 (code for $p_i$)}
\label{algo:reduction-k-IS-to-cons}
}
\end{algorithm}

The behavior of a process $p_i$ can be decomposed in three parts.
\begin{itemize}
\vspace{-0.2cm}
\item  
When it invokes  $\writesnap_k(v_i)$, 
$p_i$ first deposits its value $v_i$ in $\REG[i]$, in order all processes
can know it, and waits until  at least $(n-k)$ processes have deposited
their input value in $\REG[1..n]$
(lines~\ref{Reduction-Cons-01}-\ref{Reduction-Cons-02}). 
\vspace{-0.2cm}
\item  
Then $p_i$ proposes to the underlying consensus object $\CONS$,
the set of all the pairs
$\langle j,\REG[j] \rangle$  such that $\REG[j]\neq\bot$
(lines~\ref{Reduction-Cons-03}-\ref{Reduction-Cons-04}). 
Let us notice that this set contains at least $(n-k)$ pairs.
Hence, the consensus object returns to $p_i$ a view $view_i$, which
contains at least $(n-k)$ pairs.
\vspace{-0.2cm}
\item  Finally, $p_i$ returns a view  (of at least $(n-k)$ pairs).
\begin{itemize}
\vspace{-0.2cm}
\item 
  If $view_i$ contains its own pair $\langle i,v_i\rangle$,
  $p_i$ returns $view_i$ (line~\ref{Reduction-Cons-06}).
\vspace{-0.1cm}
\item  If $view_i$ does not contain  $\langle i,v_i\rangle$,
  $p_i$ proposes $v_i$ to the underlying immediate snapshot object
  from which it obtains a set pairs $aux_i$ (line~\ref{Reduction-Cons-07}).
  Let us notice that,
  due the properties of  the immediate snapshot object $\ISNAP$,
  $aux_i$ contains  the pair $\langle i,v_i\rangle$.
  Process $p_i$ then adds $aux_i$ to $view_i$  (line~\ref{Reduction-Cons-08})
  and returns it (line~\ref{Reduction-Cons-09}). 
\end{itemize}
\end{itemize}

\begin{theorem}
\label{theo:algo-cons-to-k-IS}
Algorithm~{\em\ref{algo:reduction-k-IS-to-cons}}  implements  {\em $k$-IS} 
 in $\CARW[0<t\leq k\leq n-1,\mbox{\em CONS}]$. 
\end{theorem}

\begin{proofT}
Proof of $k$-IS Self-inclusion.
If $p_i$ returns at line~\ref{Reduction-Cons-06},
self-inclusion follows directly from the predicate of
line~\ref{Reduction-Cons-05}.
If this predicate is not satisfied, $p_i$ invokes
the underlying immediate snapshot object $\ISNAP$ 
with the value $v_i$ it initially proposed (line~\ref{Reduction-Cons-07}). 
It then follows from the  self-inclusion property of  $\ISNAP$
that $aux_i$ contains $\langle i,v_i\rangle$, and due to
line~\ref{Reduction-Cons-08}, the set $view_i$ that is returned
at line~\ref{Reduction-Cons-09} contains $\langle i,v_i\rangle$.\\

Proof of $k$-IS Validity.
This property follows from (a) the fact that a process 
$p_i$ assigns to $\REG[i]$ the value it wants to deposit in the $k$-IS object, 
(b) this atomic variable is written at most once  
(line~\ref{Reduction-Cons-01}), and (c) the predicate  $\REG [j]\neq\bot$ 
is used at line~\ref{Reduction-Cons-03} to extract values from $\REG [1..n]$. \\

The Output size property follows from (a) the predicate of 
line~\ref{Reduction-Cons-02}, which ensures that the set $view_i$
obtained at line~\ref{Reduction-Cons-04} from the underlying
consensus  object contains at least $n-t\geq n-k$ pairs, and the fact
that a set $view_i$ cannot decrease  (line~\ref{Reduction-Cons-08}).\\

Proof of $k$-IS Containment.
Let P6 (resp., P9) the set of processes that terminate at
line~\ref{Reduction-Cons-06} (resp., \ref{Reduction-Cons-09}).
Let $view$ be the set of pairs decided by the underlying consensus object
$\CONS$ (line~\ref{Reduction-Cons-04}).
Hence, all the processes in P6 return $view$. 
Due to line~\ref{Reduction-Cons-08}, the
set $view_i$ returned by a process that  terminates at
line~\ref{Reduction-Cons-09} includes $view$. It follows that
$\forall~p_j\in$ P6, $p_i\in$ P9, we have $view_j=view\subset view_i$.

Let us now consider two processes $p_i$ and $p_j$ belonging to  P9. 
It then follows from the IS Containment property
of the underlying $\ISNAP$ object, that
we have $aux_i\subseteq aux_j$ or  $aux_j\subseteq aux_i$ 
(where the value of $aux_i$ and $aux_j$
are the ones at line~\ref{Reduction-Cons-07}).
Consequently, at line~\ref{Reduction-Cons-08} we have 
$view_i\subseteq view_j$ or  $view_j\subseteq view_i$, which completes the
proof of the $k$-IS Containment property. \\

Proof of $k$-IS Immediacy.
Let $p_i$ and $p_j$ be two processes that return $view_i$ and $view_j$,
respectively, such that $\langle i,v\rangle\in view_j$.
We have to show that $view_i \subseteq view_j$. 
Let us considering the sets P6 and P9 defined above. 
There are three cases.
\begin{itemize}
\vspace{-0.2cm}
\item Both $p_i$ and $p_j$ belong to P6.
In this case, due to line~\ref{Reduction-Cons-04}, we have $view_i=view_j$.  
\vspace{-0.2cm}
\item $p_i$ belongs to P6, while $p_j$ belong to P9.
In this case, due to line~\ref{Reduction-Cons-08},
we have $view_i\subset view_j$.  
\vspace{-0.2cm}
\item Both $p_i$ and $p_j$ belong to P9.
  In this case, due to the IS Immediacy property of $\ISNAP$
  we have (at line~\ref{Reduction-Cons-08})
  $\langle i,-\rangle \in aux_j$ $\Rightarrow$ $aux_i\subseteq aux_j$
  (and $\langle j,-\rangle \in aux_i$ $\Rightarrow$ $aux_j\subseteq aux_i$).
  Let $view$ the set of pairs returned by the consensus object
  line~\ref{Reduction-Cons-04}.
  As, due to line~\ref{Reduction-Cons-09},
  we have $view_i \leftarrow view \cup aux_i$
  and $view_j \leftarrow view \cup aux_j$, the  $k$-IS Immediacy
  property follows.   
\end{itemize} 

Proof of $k$-IS Termination.
Let $p$ be the number of processes that deposit a value in $\REG$.
As $t\leq k$, we have  $n-k \leq n-t \leq p \leq n$.
It follows that no correct process can wait forever at
line~\ref{Reduction-Cons-02}. 

The fact that no correct process blocks forever at
line~\ref{Reduction-Cons-04} and line~\ref{Reduction-Cons-07}
follows from the termination property of the underlying consensus
and immediate snapshot objects. 
\renewcommand{\toto}{theo:algo-cons-to-k-IS}
\end{proofT}


\subsection{When consensus and  $k$-IS are equivalent} 

Let us consider the right triangular matrix defined by the entries are marked
``$x$-SA'' in Table~\ref{fig:matrix-example}.
Theorem~\ref{theo:algo-cons-to-k-IS} states that it is possible to 
 to implement $k$-IS from CONS for any  entry $(t,k)$ 
belonging to this triangular matrix.  Combined with 
Corollary~\ref{coro:cartography}, we obtain the following theorem.

\begin{theorem}
\label{theo:cons-vs-t-IS}
Consensus and  {\em $k$-IS} are equivalent in 
$\CARW[0<t<n/2, t\leq k\leq (n-1)-t]$.  
\end{theorem}

\section{When Consensus is Stronger than  $k$-Immediate Snapshot}
\label{sec:cons-to-k-IS}

Section~\ref{sec:k-IS-to-x-SA}  investigated the power of
$k$-IS to implement  $x$-SA objects, namely $x$-SA can be implemented 
in $\CARW[1 \leq t \leq k < n-1, \mbox{$k$-IS}]$ where $x=\max(1,t+k-(n-2))$,
see Theorem~\ref{theo:cartography}. As we have seen, 
considering the other direction, Section~\ref{sec:equiv-to-cons} has shown 
that $k$-IS can be implemented in  $\CARW[1 \leq t \leq k < n-1,\mbox{CONS}]$
(Theorem~\ref{theo:algo-cons-to-k-IS}). 
The combination of these  results showed that Consensus and $k$-IS 
are equivalent in $\CARW[0<t=k<n/2]$ (Theorem~\ref{theo:cons-vs-t-IS}).

This section shows an upper bound on the power of $k$-IS to implement  
$x$-SA objects, namely, $k$-IS objects are not powerful enough 
to implement consensus in $\CARW[n/2 \leq t \leq k < n-1]$.

\paragraph{Preliminary: a simple lemma}
Let us remark that, as immediate snapshot objects that they
generalize, $k$-immediate snapshot objects are not linearizable.  As a
$k$-IS object $\ISNAP$ contains values from at least $(n-k)$ processes, at
least $(n-k)$ processes must have invoked the operation
$\ISNAP.\writesnap_k()$ for any invocation of $\writesnap_k()$ be able
to terminate.  It follows that there is a time $\tau$ at which $n-k$
processes have invoked $\ISNAP.\writesnap_k()$ and have not yet
returned.  We then say that these $(n-k)$ processes are ``\emph{inside}
 $\ISNAP$''. Hence the following lemma.

\begin{lemma}
\label{lemma:sim-cons}
If an invocation of $\writesnap_k()$ on a $k$-immediate snapshot
object $\ISNAP$ terminates, there is a time $\tau$ at which at least
$(n-k)$ processes are {\em inside} $\ISNAP$.
\end{lemma}

\begin{theorem}
\label{theo:simcons}
There is no algorithm implementing  consensus in 
 $\CARW[n/2\leq t\leq k <n-1,\mbox{\em{$k$-IS}}]$.
\end{theorem}

\begin{proofT}
To prove the theorem, let us first consider first the case $n=2t$.  The
proof is by contradiction.  Let us assume that $\cal A$ is a
$t$-resilient consensus algorithm for a set of processes
$\{p_1,\cdots, p_n\}$ which uses a $k$-IS object in a system where $n=2t$.
The contradiction is obtained by simulating ${\cal A}$ with two
processes $Q_0$ and $Q_1$, such that $Q_0$ and $Q_1$ solve consensus
despite the possible crash of one of them.  As there is no wait-free
consensus algorithm for 2 processes, it follows that such a consensus
algorithm ${\cal A}$ based on $t$-immediate snapshot objects cannot
exist.  The simulation is described in Algorithm~\ref{figure:simulation}.

\begin{algorithm}[ht]
\centering{\fbox{
\begin{minipage}[t]{150mm}
\footnotesize 
\renewcommand{\baselinestretch}{2.5}
\resetline
\begin{tabbing}
aaaaa\=aaa\=aaa\=aaa\=aaa\=aaa\=aaa\kill
Let $A_0$ and $A_1$ be a partition of $\{p_1,\cdots, p_n\}$:\\
$~$ $~$ $~$ $~$ $~$ $~$ 
 $|A_0|= |A_1|= t$,  $\{p_1,\cdots, p_n\}=A_0\cup A_1$, 
and $A_0 \cap A_1=\emptyset$.\\ ~\\

Code for $Q_i$  ($i\in\{0,1\}$): \\

\line{sim-01}
\>\textbf{for all} $p_j$ in $A_i$:
 initialize $v_{p_j}$ with the initial value of $Q_i$;\\

\line{sim-02}
\>\textbf{repeat forever}\\

\line{sim-03}
\>\>\textbf{for each} $p$ in $A_i$ in a round robin way {\bf do}\\

\line{sim-04}
\>\>\>\textbf{if} next operation of $p$ is $op(o,v)$ 
  (i.e. $\writesnap(v)$  on the  $k$-IS object $o$) \\

\line{sim-05}
\>\>\>\>
 \textbf{then} \= $prop_i[o] \leftarrow prop_i[o]\cup \{(p,v)\}$;\\

\line{sim-06}
\>\>\>\>\> \textbf{if} \= $\REG[i][o] = \bot$ \\

\line{sim-07}
\>\>\>\>\>\> \textbf{then} \= \textbf{if} \= $\REG[1-i][o] \neq \bot$\\

\line{sim-08}
\>\>\>\>\>\>\>\>  \textbf{then} \= 
                $\REG[i][o]\leftarrow\REG[1-i][o]\cup \{(p,v)\}$;\\

\line{sim-09}
\>\>\>\>\>\>\>\>\> 
simulation of  $op(o,v)$ for $p$ which returns $\REG[i][o]$\\

\line{sim-10}
\>\>\>\>\> \>\> \textbf{end if}\\

\line{sim-11}
\>\>\>\>\> \>\textbf{else} \> 
    $\REG[i][o]\leftarrow     \REG[i][o]  \cup \{(p,v)\}$;\\

\line{sim-12}
\>\>\>\>\>\>\>  
simulation of  $op(o,v)$ for $p$ which returns $\REG[i][o]$\\

\line{sim-13}
\>\>\>\>\> \textbf{end if}\\

\line{sim-14}
\>\>\>\> \textbf{else} \> simulate the next operation of $p$;\\

\line{sim-15}
\>\>\>\>\> \textbf{if} $p$ decides $v$ in this step \textbf{then} $Q_i$
decides $v$  \textbf{end if}\\

\line{sim-16}
\>\>\>\textbf{end if};\\

\line{sim-17}
\>\>\>\textbf{if} \= ($(|prop_i(o)|=t)~\wedge~(\REG[i][o] = \bot)$) \\

\line{sim-17-bis}
\>\>\> \>\textbf{then} 
                   $\REG[i][o] \leftarrow \ISNAP[o].\writesnap(prop_i(o))$ \\
\line{sim-18}                   
\>\>\> \textbf{end if}\\

\line{sim-19}
\>\>\textbf{end for}\\

\line{sim-20}
\> \textbf{end repeat}.
\end{tabbing}
\end{minipage}
 }
\caption{Simulation of ${\cal A}$ by  $Q_i$ ($i\in\{0,1\}$) for $n=2t$}
\label{figure:simulation}
}
\end{algorithm}

Let $A_0$ and $A_1$ be a partition of $\{p_1,\cdots, p_n\}$ such that 
$|A_0|=|A_1|=t$. $Q_0$ simulates the processes in
$A_0$, while $Q_1$ simulates the processes in $A_1$. 
In the simulation, if $Q_i$ is correct, then each simulated process in
$A_i$ executes its sequence of operations (it is consequently
correct in the simulated run).  
If $Q_i$ crashes,  its crash  entails 
(in the  simulated run) the crashes of  all the processes in $A_i$.
Note that, as at most $t$ simulated processes may crash in a simulated run, 
no process of $A_{1-i}$ crashes if all processes of $A_i$ crash.

In the following, given a simulated process $p$, 
and a $k$-IS object $o$, $op(o,v)$
denotes the invocation by $p$ of $\writesnap_k(v)$ by $p$ on the $k$-IS
object $o$.  The underlying idea of the simulation is that a
$1$-IS object accessed by $Q_0$ and $Q_1$ allows them to simulate a
$k$-IS object shared by the simulated processes $p_1$, ..., $p_n$.

The $1$-IS object   associated with 
the simulated $k$-IS object $o$, is denoted   $\ISNAP[o]$. 
Hence, in the following ``$\writesnap_k()$'' refers to an operation on 
a simulated object $o$, while $\writesnap_1()$'' refers to an operation 
issued by a simulator on a simulation object  $\ISNAP[o]$. 

In addition to the  $1$-IS objects, the 
simulator processes $Q_0$ and $Q_1$ manage the following variables. 
\begin{itemize}
\vspace{-0.2cm}
\item
 $\REG[0,1][o]$ is an array made up of two atomic read/write registers
associated with each simulated $k$-IS object $o$. 
$\REG[i][o]$ is written by $Q_i$ and read by both $Q_i$ and $Q_{1-i}$. 
It  contains (at least) the values written in $o$ by the processes 
simulated by  $Q_i$ (lines~\ref{sim-08} and~\ref{sim-11}). 
If $Q_i$ has not already simulated $\writesnap_k()$ on $o$
while $Q_{1-i}$ has, $\REG[i][o]$ is initialized to the result of the  
$\writesnap_k()$ operations on $o$ issued by the processes of $A_{1-i}$ 
simulated by $Q_{1-i}$ (lines~\ref{sim-06}-\ref{sim-08}).
\vspace{-0.2cm}
\item
$prop_i[o]$ is a local variable of $Q_i$ containing the values  written  in  
the $k$-IS object $o$ by the simulated processes in $A_i$ (line~\ref{sim-05}). 
When the next step of all the simulated processes is 
 $\writesnap_k()$ on  $o$, $Q_i$ returns the initial value of $\REG[i][o]$
(line~\ref{sim-18}). 
In the next $t$ executions of the  loop,  when $Q_i$ considers the simulated 
process $p$, this value  will be returned to $p$ (line~\ref{sim-12}) 
by  the simulation of  $\writesnap_k()$  on $o$ issued by $p$. 
\end{itemize}

The central point of the simulation lies in the way the $k$-IS objects
are simulated. For this, only when the next step of \emph{all} the
simulated processes in $A_i$ are $o.\writesnap_k()$ ($\writesnap_k()$
on the \emph{same} object $o$), the simulator $Q_i$ performs
$\writesnap_1()$ on associated $1$-IS object $\ISNAP[o]$ shared by
$Q_0$ and $Q_1$, where the values written by the processes in $A_i$ in
this $k$-IS object $o$.  The result of this invocation of 
$\writesnap_1()$  contains either all the values from all simulated 
processes, or only the values of the processes in $A_i$. Moreover, 
all processes of $Q_i$ obtain the same result, and $Q_i$ also writes 
this result value into $\REG[i,o]$ (line~\ref{sim-18}).

Let us now consider the case in which  the next step of the processes in
$A_i$ is not $\writesnap_k()$  on the same object. 
If the next step of some process $p\in A_i$ is $\writesnap_k()$
on object $o$ and no $\writesnap_k()$ on $o$ 
by processes in $A_i$  has already returned, we 
prove that there is a time $\tau$ at which all processes in $A_0$,  
or all processes $A_1$, are {\em inside} the $k$-IS object $o$. 
To this end, let us assume that there is no time  at which all processes 
in $A_i$ are inside a $k$-IS object $o$. 
By Lemma~\ref{lemma:sim-cons} there is a time $\tau$ at which a set 
of at least $k$ processes, say $C$, are inside a $k$-IS object  $o$. 
At this time, as --by assumption--  at least one process in
$A_i$ is not inside a $k$-IS object, it follows that 
at least one process of $A_{1-i}$ is inside a  $k$-IS object.
But let us then consider the run in which all
processes in $A_i$ crash (in particular all processes in $A_i$
may be considered as crashed before they invoked $\writesnap_k()$ on $o$). 
Hence for this run, $C$ contains no process in $A_i$ and, as $|C|\geq k$, 
$C$ is equal to $A_{i-1}$.

From this observation we deduce that either there is a time for which
the next operation of all $p\in A_i$ is a $\writesnap_k()$ on $o$, or
there is a time at which the next step of all processes $p\in A_{1-i}$
is a $\writesnap_k()$ on $o$.  Hence, $Q_i$ or $Q_{1-i}$ executes
$\writesnap_1()$ on $\ISNAP[o]$).  If $Q_{1-i}$ performs executes
$\writesnap_1()$ on $\ISNAP[o]$, the result for each process in
$A_{1-i}$ is the set $V$ made up of the values written by the
processes in $A_{1-i}$.  After that, $Q_i$ can read $V$ from a shared
variable, and is able to compute the result of a  $\writesnap_k()$
on $o$ (the result is $V$ union the set of values of processes 
in $A_i$ for which $Q_i$ has simulated the $\writesnap_k()$ on $o$).  
Hence, if $p\in A_i$ is stuck in the simulation on an object
$o$, either $Q_{1-i} $ eventually executes $\writesnap_1()$ on
$\ISNAP[o]$, and $Q_i$ eventually simulates $\writesnap_k()$ on $o$
for $p$, or eventually the next operation of all processes in $A_i$ is
a $\writesnap_k()$ on $o$, and $Q_i$ can compute the result returned
by these $\writesnap_k()$ on $o$.\\

To extend the result to $2t>n$,  we partition $\{p_1,\cdots,p_n\}$
in 3 sets  $A_0,  A_1, D$  such that $|A_0|=n-t$, $|A_1|=n-t$, $|D|=2t-n$.
Then,  we run the previous simulation algorithm ${\cal A}$ where all 
processes in $D$ are initially crashed, $Q_0$ simulates the set of 
processes of  $A_0$,  and $Q_1$ simulates the processes of $A_1$.
With this simulation,  $Q_0$ and $Q_1$ realizes a wait-free consensus,
which is known to be  impossible.
\renewcommand{\toto}{theo:simcons}
\end{proofT}

\begin{figure}[ht]
\centering{ \scalebox{0.35}[0.35]
{\input{fig-cartography-global.pstex_t}}
\caption{Summarizing the results}
\label{fig:cartography}
}
\end{figure}
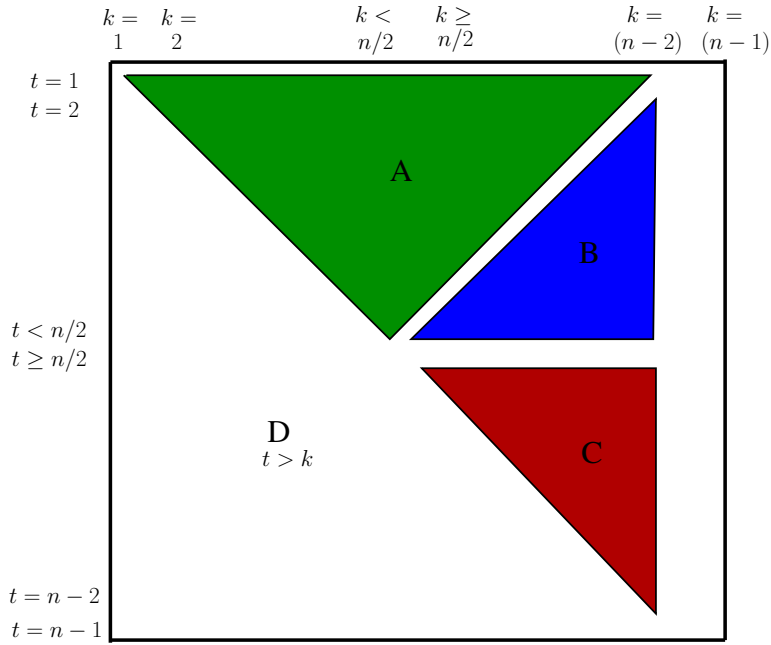

\section{Conclusion}
\label{sec:conclusion}

\paragraph{The aim and content of the paper}
The paper has first introduced the notion of a $k$-immediate snapshot ($k$-IS)
object, which generalizes the notion of immediate snapshots (IS) objects to
$t$-crash $n$-process systems 
(the IS object corresponds to the case $k=t=n-1$).
It has then shown that $k$-IS objects cannot be implemented in 
asynchronous read/write systems for $k<n-1$. 

The paper considered then the respective power of 
$k$-IS objects and $x$-set agreement objects ($x$-SA) in $t$-crash-prone 
systems. As both these family of objects are impossible to implement in 
read/write systems for $t,k<n-1$ or $x \leq t$, respectively, 
the paper strove to establish which of  $k$-IS and $x$-SA
objects are the most ``impossible to solve''. 
The main results are the following where the zones A, B, C, D, refer to 
Figure~\ref{fig:cartography}. 
\begin{itemize}

\vspace{-0.2cm}
\item Even if we have consensus objects,
it is not possible to implement $k$-IS objects in a $t$-crash system
where $t>k$ (Zone D). 
\vspace{-0.2cm}
\item
It is possible to implement $x$-SA objects, where $x=\max(1,t+k-(n-2))$, 
from  $k$-IS objects in systems where  $1\leq t \leq k <n-1$
(Zone A +  B + C). 
\vspace{-0.2cm}
\item
It is possible to implement $k$-IS objects from 
$1$-SA  objects  (consensus) in read/write systems where 
$1\leq t \leq k\leq n-1$ (Zone A +  B + C).

\vspace{-0.2cm}
\item
$1$-SA objects (consensus) and  $k$-IS objects are equivalent in 
 read/write systems where $1\leq t <n/2$ and $t\leq k \leq (n-1)-t$
(Zone A). 
\vspace{-0.2cm}
\item
It is not possible to implement $1$-SA (consensus) 
from $k$-IS objects in read/write systems when $n/2 \leq t\leq k < n-1$
(Zone C). 
 
\end{itemize}

Stated in a more operational way, these results exhibit the price of the 
synchronization hidden in $k$-IS object  (which requires that the view 
returned to a process contains at least $(n-k)$ pairs, (where a pair is
made up of a value plus the id of the process that deposited it in the 
$k$-IS object). 

More generally, the previous results establish a computability map 
relating important problems, which are impossible to solve in pure 
read/write systems.


\paragraph{Open problems}
The following problems remain to be solved to obtain a a finer relation 
linking $k$-IS and $x$-SA, when  $>1$. 
\begin{itemize}
\vspace{-0.2cm}
\item Direction ``from $k$-IS  to $x$-SA''.
Is it possible to implement $x$-SA objects,  with $1\leq x<t+k-(n-2)$
in $t$-crash $n$-process systems enriched with $k$-IS objects (Zone B)?
We conjecture that the answer to this question is $no$.
\vspace{-0.2cm}
\item Direction ``from $x$-SA to $k$-IS''.
  Given an  $x$-SA object, which $k$-IS objects can be implemented from it?
  More generally, is there a ``$k$-IS-like'' communication object
  such that $x$-SA and this  ``$k$-SA-like'' object are computationally
  equivalent   (by ``$k$-IS-like'' we mean an object possibly weaker than a
  $k$-IS object)? 
%
\end{itemize}

\section*{Acknowledgments}
This work has been partially supported by the  French  ANR  
project DESCARTES devoted to the study of abstraction layers in 
distributed computing, the UNAM-PAPIIT project IN107714, and the INRIA-UNAM
{\it \'Equipe Associ\'ee} LiDiCo (At the Limits of Distributed Computing).



\end{document}

%% file: fig-cartography-global.pstex_t
\begin{picture}(0,0)%
\includegraphics{fig-cartography-global.pstex}%
\end{picture}%
\setlength{\unitlength}{4144sp}%
\begingroup\makeatletter\ifx\SetFigFont\undefined%
\gdef\SetFigFont#1#2#3#4#5{%
  \reset@font\fontsize{#1}{#2pt}%
  \fontfamily{#3}\fontseries{#4}\fontshape{#5}%
  \selectfont}%
\fi\endgroup%
\begin{picture}(12174,10956)(-374,-11815)
\put(1396,-1636){\makebox(0,0)[lb]{\smash{{\SetFigFont{25}{30.0}{\rmdefault}{\mddefault}{\updefault}{\color[rgb]{0,0,0}$1$ }%
}}}}
\put(9856,-1636){\makebox(0,0)[lb]{\smash{{\SetFigFont{25}{30.0}{\rmdefault}{\mddefault}{\updefault}{\color[rgb]{0,0,0}$(n-2)$ }%
}}}}
\put(11341,-1636){\makebox(0,0)[lb]{\smash{{\SetFigFont{25}{30.0}{\rmdefault}{\mddefault}{\updefault}{\color[rgb]{0,0,0}$(n-1)$ }%
}}}}
\put(9271,-5281){\makebox(0,0)[lb]{\smash{{\SetFigFont{34}{40.8}{\rmdefault}{\mddefault}{\updefault}{\color[rgb]{0,0,0}B}%
}}}}
\put(6076,-3886){\makebox(0,0)[lb]{\smash{{\SetFigFont{34}{40.8}{\rmdefault}{\mddefault}{\updefault}{\color[rgb]{0,0,0}A}%
}}}}
\put(3916,-8746){\makebox(0,0)[lb]{\smash{{\SetFigFont{34}{40.8}{\rmdefault}{\mddefault}{\updefault}{\color[rgb]{0,0,0}$t>k$}%
}}}}
\put(4006,-8341){\makebox(0,0)[lb]{\smash{{\SetFigFont{34}{40.8}{\rmdefault}{\mddefault}{\updefault}{\color[rgb]{0,0,0}D}%
}}}}
\put(9316,-8701){\makebox(0,0)[lb]{\smash{{\SetFigFont{34}{40.8}{\rmdefault}{\mddefault}{\updefault}{\color[rgb]{0,0,0}C}%
}}}}
\put(  1,-2311){\makebox(0,0)[lb]{\smash{{\SetFigFont{25}{30.0}{\rmdefault}{\mddefault}{\updefault}{\color[rgb]{0,0,0}$t=1$ }%
}}}}
\put(  1,-2806){\makebox(0,0)[lb]{\smash{{\SetFigFont{25}{30.0}{\rmdefault}{\mddefault}{\updefault}{\color[rgb]{0,0,0}$t=2$ }%
}}}}
\put(-314,-6541){\makebox(0,0)[lb]{\smash{{\SetFigFont{25}{30.0}{\rmdefault}{\mddefault}{\updefault}{\color[rgb]{0,0,0}$t< n/2$ }%
}}}}
\put(-314,-7036){\makebox(0,0)[lb]{\smash{{\SetFigFont{25}{30.0}{\rmdefault}{\mddefault}{\updefault}{\color[rgb]{0,0,0}$t\geq n/2$ }%
}}}}
\put(-359,-11086){\makebox(0,0)[lb]{\smash{{\SetFigFont{25}{30.0}{\rmdefault}{\mddefault}{\updefault}{\color[rgb]{0,0,0}$t=n-2$ }%
}}}}
\put(-314,-11671){\makebox(0,0)[lb]{\smash{{\SetFigFont{25}{30.0}{\rmdefault}{\mddefault}{\updefault}{\color[rgb]{0,0,0}$t=n-1$ }%
}}}}
\put(10081,-1186){\makebox(0,0)[lb]{\smash{{\SetFigFont{25}{30.0}{\rmdefault}{\mddefault}{\updefault}{\color[rgb]{0,0,0}$k=$ }%
}}}}
\put(11431,-1186){\makebox(0,0)[lb]{\smash{{\SetFigFont{25}{30.0}{\rmdefault}{\mddefault}{\updefault}{\color[rgb]{0,0,0}$k=$ }%
}}}}
\put(5536,-1636){\makebox(0,0)[lb]{\smash{{\SetFigFont{25}{30.0}{\rmdefault}{\mddefault}{\updefault}{\color[rgb]{0,0,0}$n/2$ }%
}}}}
\put(6886,-1591){\makebox(0,0)[lb]{\smash{{\SetFigFont{25}{30.0}{\rmdefault}{\mddefault}{\updefault}{\color[rgb]{0,0,0}$n/2$ }%
}}}}
\put(2386,-1636){\makebox(0,0)[lb]{\smash{{\SetFigFont{25}{30.0}{\rmdefault}{\mddefault}{\updefault}{\color[rgb]{0,0,0}$2$ }%
}}}}
\put(2206,-1231){\makebox(0,0)[lb]{\smash{{\SetFigFont{25}{30.0}{\rmdefault}{\mddefault}{\updefault}{\color[rgb]{0,0,0}$k=$ }%
}}}}
\put(1216,-1231){\makebox(0,0)[lb]{\smash{{\SetFigFont{25}{30.0}{\rmdefault}{\mddefault}{\updefault}{\color[rgb]{0,0,0}$k=$ }%
}}}}
\put(5491,-1186){\makebox(0,0)[lb]{\smash{{\SetFigFont{25}{30.0}{\rmdefault}{\mddefault}{\updefault}{\color[rgb]{0,0,0}$k<$ }%
}}}}
\put(6841,-1186){\makebox(0,0)[lb]{\smash{{\SetFigFont{25}{30.0}{\rmdefault}{\mddefault}{\updefault}{\color[rgb]{0,0,0}$k\geq$ }%
}}}}
\end{picture}%